\documentclass[twocolumn,aps,superscriptaddress,preprintnumbers]{revtex4}

\usepackage[latin1]{inputenc}

\usepackage{graphicx}
\usepackage{graphics}
\usepackage{dcolumn}
\usepackage{bm}
\usepackage{color}
\usepackage{textcomp}
\usepackage{amsmath}
\usepackage{amssymb}

\usepackage{multirow}
\usepackage{setspace}

\usepackage{xcolor}

\usepackage{hyperref}
\hypersetup{
    colorlinks=true,       
    linkcolor=blue,        
    citecolor=blue,        
    filecolor=blue,        
    urlcolor=blue,         
    runcolor=blue
}

\newcommand{\ket}[1]{\left| #1 \right\rangle}
\newcommand{\bra}[1]{\left\langle #1 \right|}

\newcommand{\state}[2]{\left| #1 \rangle\langle #2 \right|}
\newcommand{\average}[3]{\left\langle #1 \right| #2 \left| #3 \right\rangle}

\newcommand{\omc}{\omega_{\mathrm{c}}}
\newcommand{\ad}{\hat{a}^{\dagger}}
\newcommand{\as}{\hat{a}}

\newcommand{\USACH}{Department of Physics, Universidad de Santiago de Chile, Avenida Ecuador 3493, Santiago, Chile.}

\newcommand{\MIRO}{Millennium Institute for Research in Optics (MIRO), Concepci\'on, Chile}

\begin{document}

\title{The Shape of the Electric Dipole Function Determines the Sub-Picosecond Dynamics of Anharmonic Vibrational Polaritons}

\author{Johan F. Triana}
\affiliation{\USACH}

\author{Federico J. Hern\'andez}
\affiliation{\USACH}
\affiliation{Department of Chemistry, School of Biological and Chemical Sciences, Queen Mary University of London, London, UK}

\author{Felipe Herrera}
\affiliation{\USACH}
\affiliation{\MIRO}
\email{felipe.herrera.u@usach.cl}

\date{\today}

\begin{abstract}
Vibrational strong coupling has emerged as a promising route for manipulating the reactivity of molecules inside infrared cavities. Growing  evidence in this direction suggests that conventional treatments of light-matter interaction are insufficient to describe the experimental results. We develop a fully quantum mechanical methodology to describe the static properties and the dynamics of a single anharmonic vibrational mode interacting with a quantized infrared cavity field in the strong and ultrastrong coupling regimes. By comparing multiconfiguration time-dependent Hartree (MCTDH) simulations for a Morse oscillator in a cavity, with an equivalent formulation of the problem in Hilbert space, we describe for the first time the essential role of permanent dipole moments in the femtosecond dynamics of vibrational polariton wavepackets. We show that depending on the shape of the electric dipole function $d_e(q)$ along the vibrational mode coordinate $q$, molecules can be classified into three general families. For molecules that are polar and have a positive slope of the dipole function at equilibrium, we show that an initial diabatic light-matter product state without vibrational or cavity excitations can evolve into a polariton wavepacket with a large number of intracavity photons, for interaction strengths at the onset of ultrastrong coupling. This build up of intracavity photon amplitude is accompanied by an effective {\it lengthening} of the vibrational mode of nearly $10\%$, comparable with a laser-induced vibrational excitation in free space. In contrast, molecules that are also polar at equilibrium but have a negative slope of the dipole function,  experience an effective mode {\it shortening}  under equivalent  coupling conditions. Our model predictions are numerically validated using realistic {\it ab-initio} potentials and dipole functions for HF and CO$_2$ molecules in their ground electronic states. We finally propose a non-adiabatic state preparation scheme to generate vibrational polaritons using nanoscale infrared antennas and UV-vis photochemistry or electron tunneling, to enable the far-field detection of spontaneously generated infrared quantum light.
\end{abstract}


\maketitle


\section{Introduction}

The recent experimental demonstrations of strong and ultrastrong light-matter interaction with molecules and molecular materials in infrared cavities \cite{Long2015,Kapon2017,Muallem2016,Simpkins2015,Thomas2016,Vergauwe2016,Chervy2018,George2015,George2016,Hertzog2017,Shalabney2015coherent,Shalabney2015raman,Dunkelberger2016,Xiang2018,Ahn2018,Dunkelberger2018,Thomas2019,Dunkelberger2019}, has stimulated intense theoretical efforts for understanding the microscopic properties of hybrid photon-vibration states from a quantum mechanical perspective \cite{Hernandez2019}. Motivated by pioneering measurements in liquid-phase Fabry-Perot cavities \cite{Thomas2016,Thomas2019,Vergauwe2019}, theoretical studies have focused on the analysis of potential mechanisms that would enable the modification of chemical reactivity in the ground electronic state, under conditions of vibrational strong coupling \cite{Martinez-Martinez2018a,Campos-Gonzalez-Angulo2019,Galego2019}. Another current theoretical focus is the study of linear and nonlinear spectroscopic signals of strongly coupled infrared cavities \cite{delPino2015raman,Saurabh2016,Strashko2016,Ribeiro2018,Hernandez2019}. 

Several models with varying degrees of complexity have been used to study the static and dynamical properties of vibrational polaritons, which are the hybrid light-matter states that emerge in infrared cavities under strong coupling \cite{Herrera2020perspective}. In one of the earliest approaches to this problem \cite{delPino2015}, molecular vibrations were treated as two-level systems with an energy gap given by the fundamental vibration frequency. This effective vibrational {\it qubit} was  coupled to a quantized simple harmonic oscillator representing  a single-mode cavity field, within the rotating-wave approximation. For an ensemble of molecules, this approach  corresponds to the Tavis-Cummings model of cavity quantum electrodynamics (QED \cite{Kirton2019}). Improving over this restricted model, intramolecular  vibrations within an electronic state are considered as quantum harmonic oscillators in Ref. \cite{Strashko2016}. The light-matter coupling model was also extended to include counter-rotating and self-energy terms that are commonly considered under conditions of ultrastrong coupling \cite{Kockum2019,Forn-Diaz2018}, broadly defined as the regime in which the light-matter interaction energy is comparable with the vibrational and cavity frequencies. 

The anharmonicity of molecular vibrations was first taken into account in Ref. \cite{Saurabh2016},  to model the pump-probe spectrum of strongly coupled infrared cavities \cite{Dunkelberger2018}. The authors also treated  vibrations as  quantum harmonic  oscillators, but supplemented with a weak anharmonic correction to the potential. The latter is introduced to capture the spectral anharmonicity of real vibrations (e.g., the spectrum is not equally spaced). Later in Ref. \cite{Ribeiro2018}, this perturbative approach was further extended to include the effect of electrostatic anharmonicity (e.g., the dipole moment function is not symmetric relative to displacements from equilibrium). Light-matter interaction with the cavity field was treated within the rotating-wave approximation in both works. 

Perturbative anharmonicity models are best suited to describe the spectroscopy of fundamental and overtone transitions starting from the ground level ($\nu=0$) up to the second vibrational state ($\nu=2$) \cite{Demtroder-book}. This limited range of applicability is not problematic when discussing free-space infrared spectroscopy, as the light-matter coupling can be safely regarded to be perturbative outside an infrared cavity. However, under conditions of strong and ultrastrong light-matter interaction inside a resonant cavity, one expects the contribution of higher vibrational levels ($\nu\geq 3$) to be non-negligible. 

The first consistent approach that takes into account the entire bound state spectrum and electric dipole function of an electronic potential energy curve was developed in Ref. \cite{Hernandez2019}, for an individual vibrating molecule in an infrared cavity vacuum, giving rise to the multi-level quantum Rabi model (MLQR). The model can be derived  from a valid electric-dipole formulation of light-matter interaction that does not ignore dipole self-energy terms \cite{Andrews2018}, by projecting  the system Hamiltonian into the vibrational energy basis. For a given electronic potential curve and electric dipole function, the MLQR model can be used to understand both material and photonic properties of vibrational polaritons both in the strong and ultrastrong coupling regimes. 

Only molecules without electric dipole moment at equilibrium were considered in Ref. \cite{Hernandez2019}. Moreover, only off-diagonal dipole matrix elements in the vibrational energy basis (transition dipoles) were taken into account in the light-matter interaction process. We now significantly expand the MLQR model to include the contribution of diagonal elements of the dipole matrix in the vibrational basis (permanent dipoles), and compare the resulting polariton physics of molecular vibrations that are polar at equilibrium (e.g., CO) with those that are non-polar at equilbrium (e.g. CO$_2$). We show that the extended MLQR model is equivalent to a formulation of the problem in the coordinate representation for both material and photonic variables \cite{Triana2018,Triana2019},  which is believed to be numerically exact. This equivalence has been widely ignored in the literature. We are particularly interested in the evolution of vibrational polariton wavepackets and the corresponding dynamics of simple material and photonic observables such as the mode length and the photon number. 

We perform a systematic comparison between polar and nonpolar molecules under several conditions of non-perturbative light-matter interaction, correlating the entire shape electric dipole function along the nuclear coordinate with the resulting vibrational polaritons dynamics. Radiative and non-radiative dissipation is not taken into account and thus our results focus on the sub-picosecond dynamics of the light-matter system. Among other results, we predict that for a specific class of polar molecules, the light-matter system can evolve from a diabatic product state with a definite number of vibrational and cavity excitations (possibly vacuum), into a polariton wavepacket with a mean intracavity photon number that could be reliably measured using current detector technology. Our work can thus stimulate further experimental work on nanoscale vibrational strong coupling \cite{Luxmoore2014,Metzger2019,Ojambati2019}. 

In the rest of the article, we first discuss the theoretical foundations of our model (Sec. \ref{sec:framework}) and the details of our methods (Sec. \ref{sec:methods}). We then describe the results obtained for the static and dynamical properties of vibrational polaritons that emerge under the various material and photonic conditions considered (Sec. \ref{sec:results}). We finally conclude with a  discussion of the fundamental physical principles that support our numerical results, and propose a quantum state preparation method that could be used to test our predictions in nanoscale photonic devices (Sec. \ref{sec:discussion}).


\section{Theoretical Framework}
\label{sec:framework}

Following Refs. \cite{Kowalewski2016,Triana2018,Triana2019}, we model  an individual molecular vibration coupled to a single quantized electromagnetic mode using a Hamiltonian that in the electric dipole approximation for the light-matter interaction can be written in the coordinate picture as 
\begin{eqnarray}\label{eq:Htotal coordinate}
\nonumber\hat{\mathcal{H}}&=&  \left( -\frac{1}{2\mu}\frac{\partial^{2}}{\partial q^{2}} +  {V}(q) \right) + \left(-\frac{1}{2}\frac{\partial^{2}}{\partial \hat{x}^{2}} + \frac{1}{2}\omc^{2}\hat{x}^{2} \right) \\
& & + \sqrt{2}\omc\mathcal{E}_{0}\hat{x}\hat{d}(q). 
\end{eqnarray}
The first term describes the vibrational motion of interest, characterized by a potential energy curve (PEC) $V(q)$ along the normal mode coordinate $q$ that has a reduced mass $\mu$. In this work, the PEC is assumed to have a single equilibrium configuration at $q=q_e$ and a well-defined dissociation energy in free space. This behavior can describe realistic stretching vibrational modes in diatomic and polyatomic molecules \cite{Demtroder-book}.  

The second term in Eq. (\ref{eq:Htotal coordinate}) describes the energy of a single cavity field mode of frequency $\omega_c$ and quadrature operator $\hat x$. The third term corresponds to light-matter interaction in multipolar form \cite{Andrews2018}, truncated to the electric dipole approximation. $\mathcal{E}_0$ denotes the square-root amplitude of the vacuum  fluctuations at $\omega_c$ and $\hat d(q)$ is the electric dipole function along the vibrational coordinate. Atomic units are used throughout, unless otherwise specified. 

The evolution of an arbitrary intracavity light-matter state $\ket{\psi(t)}$ is obtained from $\hat{\mathcal{H}}$ in Eq. (\ref{eq:Htotal coordinate}) by solving $i(d/dt) \ket{\psi (t)}=\hat{\mathcal{H}}\ket{\psi(t)}$ with the appropriate propagator. As described in more detail below, we propagate intracavity wavefunctions by representing  $\ket{\psi(t)}$ in coordinate space representation and in Hilbert space. In the coordinate representation, we discretize both $\hat{\mathcal{H}}$ and $\ket{\psi(t)}$ along both nuclear and cavity coordinates $(q,x)$ and propagate an initial state using the multi-configurational time-dependent Hartree method (MCTDH \cite{mctdhpaper,mctdhpack}). In the Hilbert space representation, time-evolution is carried out by first projecting Eq. \ref{eq:Htotal coordinate} into a set of energy eigenstates of the nuclear potential $V(q)$ to give 
\begin{equation}\label{eq:H projection}
\hat{\mathcal{H}}'=\hat {\Pi}_{\rm vib}\,\hat{\mathcal{H}}\,\hat{\Pi}_{\rm vib}^\dagger
\end{equation}
where $\hat{\Pi}_{\rm vib}= \sum_{\nu=0}^{\nu_{\rm max}}\ket{\nu}\bra{\nu}$ is a projector operator into the vibrational energy basis $\ket{\nu}$, with $\nu$ being the vibrational quantum number. Clearly if the energy basis set is complete, i.e., $\nu_{\rm max}=\infty$, then we have $\hat{\mathcal{H}}' \equiv \hat{\mathcal{H}} $ and  evolving a system state in the coordinate or Hilbert space representation must give {\it exactly} equivalent results. In practice, the projection operator $\hat{\Pi}_{\rm vib}$ can only reliably be truncated up to a cutoff energy eigenstate with $\nu_{\rm max}$ finite. We demonstrate below that despite this practical limitation, it is possible to find a value of $\nu_{\rm max}$ that gives numerically equivalent results when propagating vibrational polariton wavepackets either in the coordinate representation or in the Hilbert space representation.

As we discuss in detail below, we are interested in the role of diagonal and off-diagonal elements of the dipole moment matrix in the vibrational eigenbasis $\bra{\nu'}{d}(q)\ket{\nu}$, on the properties of vibrational polaritons. In order to have controllable access to this information, we partition the vibrationally-projected Hamiltonian $\hat{\mathcal{H}'}$ in Eq. (\ref{eq:H projection}) as $\hat{\mathcal{H}'}= \hat{\mathcal{H}_1}+\hat{\mathcal{H}_2}$, where 
\begin{eqnarray}\label{eq:H1}
\hat{\mathcal{H}_1}&=&\omc\ad\as + \sum_{\nu}\omega_{\nu}\state{\nu}{\nu} \\
    & +& \sum_{\nu}\sum_{\nu'> \nu} g_{\nu'\nu}(\state{\nu'}{\nu}+\state{\nu}{\nu'})(\ad+\as),\nonumber
\end{eqnarray}
and 
\begin{eqnarray}\label{eq:H2}
\hat{\mathcal{H}}_2&=&\sum_{\nu} g_{\nu}\state{\nu}{\nu}(\ad+\as),
\end{eqnarray}
where summations over $\nu$ are implicitly assumed to be carried out up to $\nu_{\rm max}$. 

The term $\hat{\mathcal{H}_1}$ in Eq. (\ref{eq:H1}) is the multi-level quantum Rabi model (MLQR), developed in Ref. \cite{Hernandez2019}, in which light-matter coupling parameters are determined only by transition dipole moments through the state-dependent Rabi frequencies
\begin{equation}\label{eq:g_nunuprime}
g_{\nu'\nu} =  \mathcal{E}_{0}\average{\nu'}{\hat{d}(q)}{\nu}, 
\end{equation}
for $\nu\neq \nu'$. Cavity field variables are described in terms of the cavity annihilation operator $\hat a$.  For a two-level system, i.e., $\nu_{\max}=1$, Eq. (\ref{eq:H1}) would reduce to the quantum Rabi model for a qubit in ultrastrong coupling \cite{Werlang08,Braak11,Wolf13}, provided that the energy eigenstates $\ket{\nu}$ have a well-defined parity. 

Noting that the intracavity electric field $\hat{\mathbf{E}}$ is proportional to $\mathcal{E}_0(\ad+\as)$ \cite{Herrera2017-PRA}, the term $\hat{\mathcal{H}}_2$ in Eq. (\ref{eq:H2}) can be interpreted as the contribution to $\mathbf{d}\cdot \mathbf{E}$ from the permanent dipole moment of each vibrational level $\nu$. The corresponding diagonal coupling strength is given by 
\begin{equation}\label{eq:g_nu}
g_{\nu} =  \mathcal{E}_{0}\average{\nu}{\hat{d}(q)}{\nu}.
\end{equation}

By partitioning Eq. (\ref{eq:H projection}) into contributions that involve either off-diagonal ($\hat{\mathcal{H}_1}$) or diagonal electric dipole matrix elements ($\hat{\mathcal{H}_2}$), we are able to isolate their contribution to the light-matter system dynamics. In particular, we compare below the evolution of intracavity system observables  without the contribution of permanent dipole moments, e.g., by setting $\hat{\mathcal{H}_2}=0$, with the evolution obtained by propagating the system using the entire electric dipole function in coordinate space, and the full dipole matrix in the vibrational eigenbasis.


The dependence on the vibrational quantum numbers of the diagonal ($g_{\nu}$) and off-diagonal ($g_{\nu'\nu}$)  coupling parameters are directly related to the potential energy curve $V(q)$. Consider a potential that has {\it even} parity relative to the equilibrium mode length $q_e$, i.e., $V(q)$ is invariant under the transformation $q\rightarrow -q$ and $q_e=-q_e$. This is the case for the harmonic potential $V(q)=\omega_0(q-q_e)^2/2$, whose vibrational eigenstates are also eigenstates of parity.

Selection rules for electric dipole matrix elements can be derived by expanding the dipole function $d(q)$ near $q_e$ up to second order as 
\begin{equation}\label{eq:dipole expansion}
d(q)\approx d_e + c_1(q-q_e)+ c_2(q-q_e)^2, 
\end{equation}
where $d_e$ is the electric dipole moment at the equilibrium configuration $q_e$, $c_1$ is proportional to the slope of the dipole function at equilibrium and $c_2$ to its curvature. The absolute magnitude dipole parameters $(|d_e|,|c_1|,|c_2|)$ can be inferred from the strengths of fundamental and overtone absorption peaks in stationary infrared spectroscopy \cite{Vasan1983}. Since the absorption line strengths are proportional to the square of the transition dipole moments, the signs of the expansion parameters cannot be resolved using linear spectroscopy. For ideal harmonic vibrations, the expansion in Eq. (\ref{eq:dipole expansion}) gives the usual selection rules $\Delta \nu=\pm1, \pm2$ for transition dipole moments. Diagonal elements (permanent dipoles) are only weakly dependent on the vibrational quantum number $\nu$ through the quadratic term in the expansion, and are thus primarily determined given by $\langle \nu|\hat d(q)|\nu\rangle \approx d_e$  for small $\nu$.  

For realistic anharmonic molecular vibrations, the nuclear potential $V(q)$ in general is not invariant under the transformation $q\rightarrow -q$ and $q_e=-q_e$. Anharmonic vibrational eigenstates $\ket{\nu}$ thus do not have a well-defined parity. This changes the structure of the dipole matrix elements  $\langle \nu'|\hat d(q)|\nu\rangle$ in comparison with the case of harmonic vibrations.  In general, for anharmonic vibrations there are no selection rules for transition dipole moments. Permanent dipole moments have a stronger dependence with the vibrational quantum number in comparison with harmonic vibrations, as the contribution of the linear term in Eq. (\ref{eq:dipole expansion}) is not forbidden, i.e, $\langle \nu|\hat d(q)|\nu\rangle\approx d_e+c_1\langle \nu|(q-q_e)|\nu \rangle$ for small $\nu$.

In this work, we consider the intracavity dynamics of an individual anharmonic vibration described by a Morse potential of the form
\begin{equation}\label{eq:Morse potential}
V(q) = D_e\left(1-{\rm exp}[-a(q-q_e)]\right)^2,
\end{equation}
where $D_e$ is the potential depth that defines the dissociation energy and $a$ is a parameter that contributes to the anharmonicity of the cavity-free vibrational spectrum. The nuclear Schr\"{o}dinger equation with a Morse potential can be solved analytically in terms of associated Laguerre polynomials \cite{Vasan1983,Morse1929}. By comparing the exact expression for the vibrational energies $E_\nu$ with the Dunham expansion \cite{Demtroder-book}, the vibrational spectrum up to second order in $\nu$ can be written as
\begin{equation}
E_\nu \approx -D_e +\omega_0(\nu+1/2)-\omega_0\chi_e(\nu+1/2)^2,
\end{equation}
where $\omega_0\equiv[\hbar]a\sqrt{2D_e/\mu}$ is the fundamental vibration frequency in the harmonic approximation, and $\chi_e\equiv[\hbar^2]a^2/2\mu$ is the spectral anharmonicity parameter. 
 
 Having defined a model for the nuclear potential and a structure for the light-matter coupling Hamiltonian both in the coordinate space [Eq. (\ref{eq:Htotal coordinate})] and Hilbert space  [Eqs. (\ref{eq:H1})-(\ref{eq:H2})], we  proceed to compute the dynamics of selected intracavity light-matter wavepackets, which in general can be written as 
 \begin{equation}\label{eq:general wavepacket}
 \ket{\psi(t)} = \sum_\nu\sum_{n_c} \beta_{\nu\,n_c}(t)\ket{\nu}\ket{n_c},
 \end{equation}
where $\beta_{\nu\,n_c}(t)$ are time-dependent wavepacket coefficient in the diabatic product basis $\ket{\nu}\ket{n_c}$, where $\ket{n_c}$ is an eigenstate of the photon number operator $\hat a^\dagger \hat a$ with eigenvalue $n_c$ (Fock state). As discussed in more detail below, a coordinate space analogue of Eq. (\ref{eq:general wavepacket}) can also be defined. 

In order to gain physical intuition about the fate of the nuclear motion and the cavity field under conditions of strong and ultrastrong light-matter coupling, we focus on the short-time dynamics of the mean mode length $\langle \psi(t)|\hat q|\psi(t) \rangle $ and the mean intracavity photon number $\langle \psi(t)|\hat a^\dagger \hat a|\psi(t)\rangle $ as a function of the light-matter coupling strength and the shape of the dipole function $d(q)$. We a devote special attention to the latter, as one would expect that for strong enough light-matter interaction, the spectral observables should depend not only on the {\it magnitudes} of the dipole expansion parameters in Eq. (\ref{eq:dipole expansion})--as is the case in cavity-free infrared absorption spectroscopy--but also on their {\it signs}.  

Based on this motivation, we classify anharmonic vibrations into modes without electric dipole moment at equilibrium ($d_e=0$), and modes that are polar at equilibrium ($|d_e|>0$). Examples of non-polar modes at equilibrium are the asymmetric CO stretching modes of CO$_2$ and Fe(CO)$_5$ molecules. These vibrational modes acquire a finite electric dipole moments away from equilibrium. 

Molecular vibrations that are polar at equilibrium are further sub-divided into those whose dipole function has a positive slope at $q_e$, [$c_1>0$ in Eq. (\ref{eq:dipole expansion})], and those that have a negative slope. We refer to the former as ``dipole-right" species, as the electric dipole has a maximum for mode lengths beyond $q_e$. Conversely, we refer to species with $c_1<0$ at $q_e$ as ``dipole-left" systems, which have a maximum dipole moment at shorter mode lengths.

\section{Methods}
\label{sec:methods}

\subsection{Morse and Dipole Function Parameters}

We consider a model anharmonic oscillator described by a Morse potential [Eq. (\ref{eq:Morse potential})] with parameters in atomic units  given by $D_{\mathrm{e}}=0.23$ a.u., $q_{\mathrm{e}}=4.0$ a.u., $\alpha=1.4465$ a.u. The reduced mass of the vibrational mode is $\mu=1.43764$ amu. The same Morse potential $V(q)$  is used for the system Hamiltonian represented in coordinate space and Hilbert space. For our chosen parameters, the potential has 24 vibrational bound states, and the fundamental vibration period is $2\pi/\omega_{10}=8.27$ fs, where $\omega_{10}$ is the frequency of the $0\rightarrow 1$  transition.

We numerically compute the vibrational energies $\omega_\nu$ and eigenstates $\ket{\nu}$ of the potential using a discrete variable representation (DVR) method with a uniform grid and Fourier basis functions \cite{Colbert1992}. We use up to $N_{q}=721$ grid points over the interval $2.5<q< 20.5$ (a.u) along the nuclear coordinate. We can construct a quasi-complete nuclear basis $\ket{\nu}$ with up to $\nu_{\rm max}\sim 700$, which includes states above the dissociation threshold. However, in   most static and dynamical calculations considered, converged results up to a very low tolerance can be obtained with $\nu_{\rm max}\sim 20-80$, depending on the coupling strength and molecular species. 

For the definition of the electric dipole function along the nuclear coordinate, we follow Ref. \cite{Hernandez2019} and consider the universal model  
\begin{equation} \label{eq:dipole function}
{d}(q)=d_{0}(q-c_{0})e^{-(q-q_{0})^{2}/2\sigma^{2}},
\end{equation}
where the set of parameters ($d_0, c_0, q_0, \sigma$) can be chosen such that the $d(q)$ can equally well describe the qualitative behavior of IR-active molecular species that are polar or non-polar at equilibrium. As is mentioned above, we compare three types of species according to the form of their dipole function $d(q)$: non-polar molecules for which $d(q_e)=0$, polar molecules with $|d(q_e)|>0$ and $(d/dq)[d(q)]|_{q_e}>0$, and finally polar molecules with  $|d(q_e)|>0$ and $(d/dq)[d(q)]|_{q_e}<0$. In what follows, we respectively denote  these cases as non-polar (NP), polar-right (PR), and polar-left (PL). The set of model function parameters used in this work are given in Table \ref{tab:dipole params}, in atomic units. In the last row, we also include the value of the transition dipole moment $d_{10}=\bra{1}d(q)\ket{0}$. 

\begin{table}[h]
    \centering
    \begin{tabular}{c c c c}\hline
             & $d_{\mathrm{PL}}(q)$  &  $d_{\mathrm{PR}}(q)$ & $d_{\mathrm{NP}}(q)$ \\
    \hline\hline
    $c_{0}$   &  1.8  &  2.7   &  4.0  \\
    $q_{0}$   &  3.6  &  4.5   &  4.0  \\
    $\sigma$  &  0.3  &  0.584 &  0.6  \\
    $d_{0}$ &  1.0  &  1.0   &  2.0  \\
     $d_{10}$ & -0.32 & 0.20 & 0.19 \\     \hline
    \end{tabular}
    \caption{Dipole moment function parameters (in atomic units) for polar-left (PL), polar-right (PR) and non-polar (NP) molecular species. The $0\rightarrow 1$ transition dipole moment ($d_{10}$) is also given.}
    \label{tab:dipole params}
\end{table}

\begin{figure}[t]
    \includegraphics[width=0.5\textwidth]{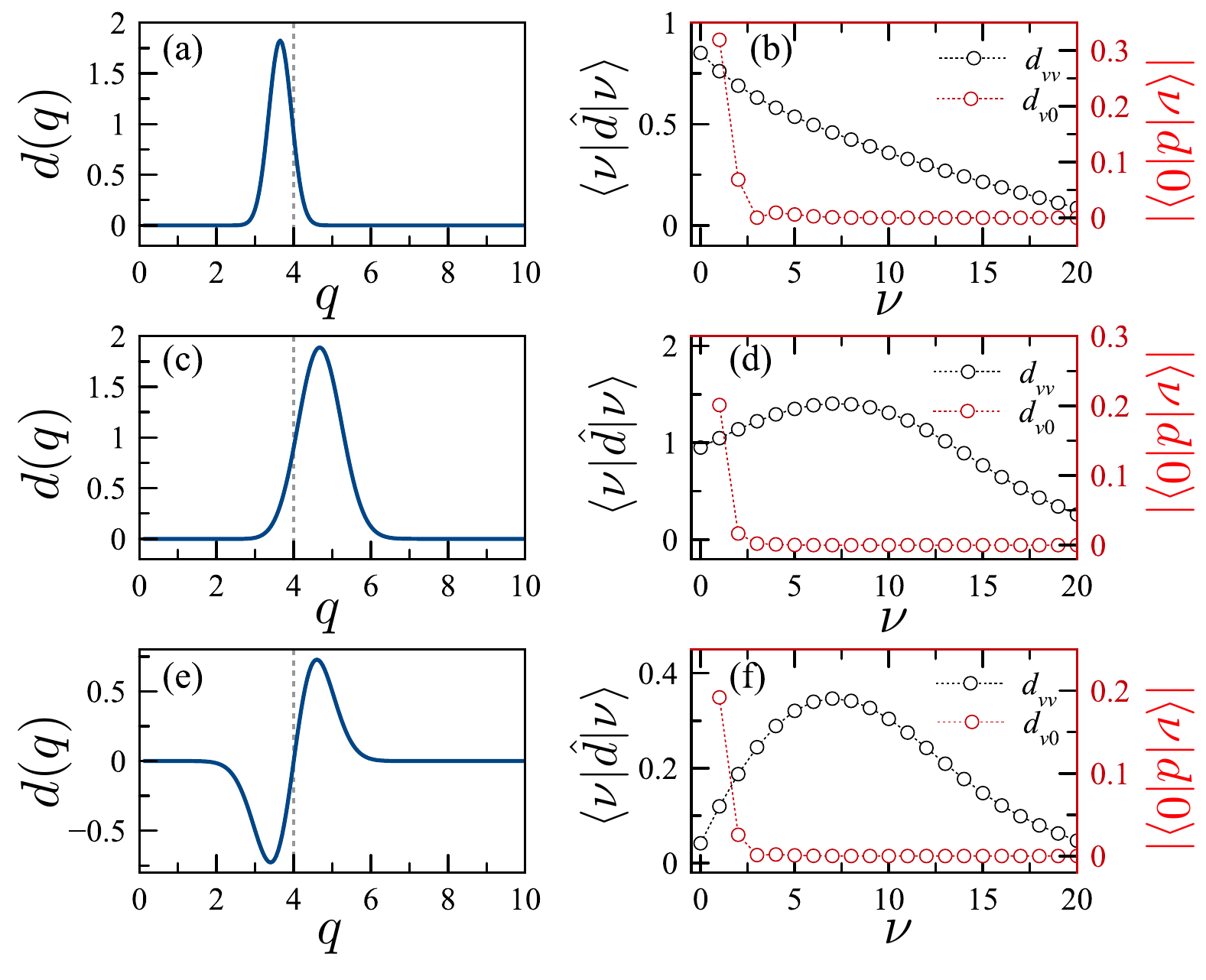}
    \caption{{\bf Dipolar properties of molecular vibrations}. (Left panels) Dipole moment function $d(q)$ (in atomic units) for molecules that are polar and non-polar at equilibrium. (Right panels) Vibrationally-averaged permanent dipole moments $\bra{\nu} d(q)\ket{\nu} $ (left axes) and transition dipole moments $|\bra{\nu} d(q)\ket{\nu}|$ for $0\rightarrow 1$ vibrational transitions (right axes). In all panels, the Morse potential parameters used are $D_e=0.23$ a.u., $q_e=4.0$ a.u., $a=1.4465$ a.u., and $\mu=1.43$ amu.}
    \label{fig:dipoles}
\end{figure}

In Fig. \ref{fig:dipoles} we plot the  dipole functions $d(q)$, permanent dipole moments $d_{\nu}$ and transition dipole moments $d_{\nu 0}$ for the first 20 bound states of each of the three types of molecular species parametrized in Table \ref{tab:dipole params}. The square of the transition dipoles $d_{\nu0}$ are proportional to the oscillator strength of the vibrational transition $0\rightarrow \nu$ in linear infrared absorption. The three model functions considered qualitatively reproduce the typical behavior in infrared absorption spectroscopy, which feature a strong fundamental peak ($0\rightarrow 1$) with weaker overtones ($0\rightarrow 2,3,\ldots$). The permanent dipole moments also behave as expected, with their magnitude decreasing rapidly as $\nu$ approaches the continuum.

\subsection{Cavity Field Parameters}

In all our calculations the infrared cavity frequency $\omega_c$ is set to be on exact resonance with the fundamental vibrational frequency $\omega_{10}$, associated with the transition $0\rightarrow 1$. The vacuum field amplitude $\mathcal{E}_{0}$ is considered as a tunable parameter, simulating the fact that in real cavities the magnitude of the light-matter coupling strength can be tuned by changing the intracavity molecular density, for a fixed cavity geometry and material composition \cite{Simpkins2015}.  For consistency between our coordinate space and Hilbert space calculations, throughout this work we parametrize the light-matter coupling strength by the dimensionless parameter 
\begin{equation}
\lambda_g \equiv \frac{\average{1}{\hat{d}(q)}{0}}{\omega_c} \,\mathcal{E}_{0}.
\end{equation}
This definition of $\lambda_g$ corresponds to the light-matter coupling ratio $g/\omega_c$ used in the ultrastrong coupling literature \cite{Kockum2019,Forn-Diaz2018}, if we define $d_{10}\mathcal{E}_0\equiv g$.
 
The dimensionality of the cavity Hilbert space is set to ensure convergence of static and dynamical calculations. For the values of $\lambda_g$ considered below, converged results were obtained by including Fock states $\ket{n_c}$ with up to $n_c\sim 80$ photons. Convergence of the calculations in the coordinate space representation is discussed below.

\subsection{Polariton Wavepacket Propagation}
\label{sec:propagation methods}

We are interested in the dynamics of expectation values of the form $\langle \psi(t)|\hat O|\psi(t)\rangle$, where $\hat O$ is any molecular or photonic observable of experimental interest such as the photon number operator ($ \hat a^\dagger \hat a$), the mode distance operator ($\hat q$), or the vibrational energy operator. The system state $\ket{\psi(t)}$ is obtained by propagating numerically an initial light-matter wavepacket $\ket{\psi_0}$ with a unitary propagator $\hat U(t)$, i.e., $\ket{\psi(t)}=\hat U(t)\ket{\psi_0}$. The propagator and the wavefunction can be accurately represented in both the Hilbert and coordinate space representations, as described in what follows.

\subsubsection{Eigenphase Evolution in Hilbert Space}

State evolution in Hilbert space is carried out by projecting the time-evolution operator $\hat U(t)={\rm exp}[-i\hat{\mathcal{H}}'t]$ into a truncated energy basis of the system Hamiltonian $\hat{\mathcal{H}'}$. For calculations that only take into account the contribution of transition dipole moments in the light-matter interaction, we set $\hat{\mathcal{H}}_2=0$ [Eq. (\ref{eq:H2})]. Energy eigenstates $\ket{\epsilon_j}$ satisfy $\hat{\mathcal{H}}'\ket{\epsilon_j}=\epsilon_j\ket{\epsilon_j}$, where $j$ labels discrete and quasi-discrete polariton energy levels. Physically, energy eigenstates correspond to anharmonic vibrational polaritons \cite{Hernandez2019}. 

We project an arbitrary initial light-matter state into the polariton basis as $\ket{\psi_0}=\sum_{j} \langle \epsilon_j |\psi_0\rangle \ket{\epsilon_j}$. In this basis, initial states evolve into polariton wavepackets that undergo  trivial phase evolution to read
\begin{equation}    \label{eq:Hilbert evolution}
    \ket{\psi(t)}=\sum_{j=0}^{j_{\rm max}}\;\langle \epsilon_j |\psi_0\rangle   \ket{\epsilon_j}\,e^{-\mathrm{i}\epsilon_j t},
\end{equation}
where the summation  includes the absolute ground state ($j=0$) and runs up to a spectral cut-off ($j_{\rm max}$), chosen such that the observables of interest $\langle \psi(t)|\hat O|\psi(t)\rangle$ are converged over the entire evolution time. In this approach, convergence of the polariton spectrum is essential for obtaining an accurate wavepacket dynamics. 

\subsubsection{MCTDH Evolution in Coordinate Space}

In the coordinate space representation, the Hamiltonian $\hat{\mathcal{H}}(x,q)$ in Eq. (\ref{eq:Htotal coordinate}) is treated as a two-dimensional potential energy surface (2D-PES) with respect to the nuclear and photonic coordinates, with the corresponding kinetic energy terms \cite{kowalewski2016cavity,Kowalewski2016,Triana2018,Csehi2019}. The light-matter interaction term proportional to $x \times d(q)$ is regarded as non-adiabatic coupling between the anharmonic nuclear potential $V(q)$ and harmonic photonic potential $V(x)\equiv \omega_c^2x^2/2$. 

In this 2D-PES representation, the evolution of an arbitrary light-matter wavepacket $\psi_0(x,q)$ is accurately computed using the multi-configurational time-dependent Hartree method (MCTDH, \cite{mctdhpaper,mctdhbook}), as implemented in Ref. \cite{mctdhpack}. By projecting a wavepacket $\psi(x,q,t)$ into the polariton eigenbasis, static properties such as the polariton spectrum ($\epsilon_j$) may also be obtained, as described below.

Static and dynamics calculations are carried out using  DVR with a sine primitive basis for the molecular coordinate $q$, in a grid of $N_{q}=721$ points in the range $2.5 \leq q\leq 20.5$ a.u. For the photonic coordinate $x$, we use harmonic oscillator primitive basis functions  on a dimensionless DVR grid within the range ($-90,90$). The number of photonic grid points $N_x$ is chosen such that all the Fock states needed for convergence can be properly described. 

\begin{figure*}[t]
    \includegraphics[width=0.9\textwidth]{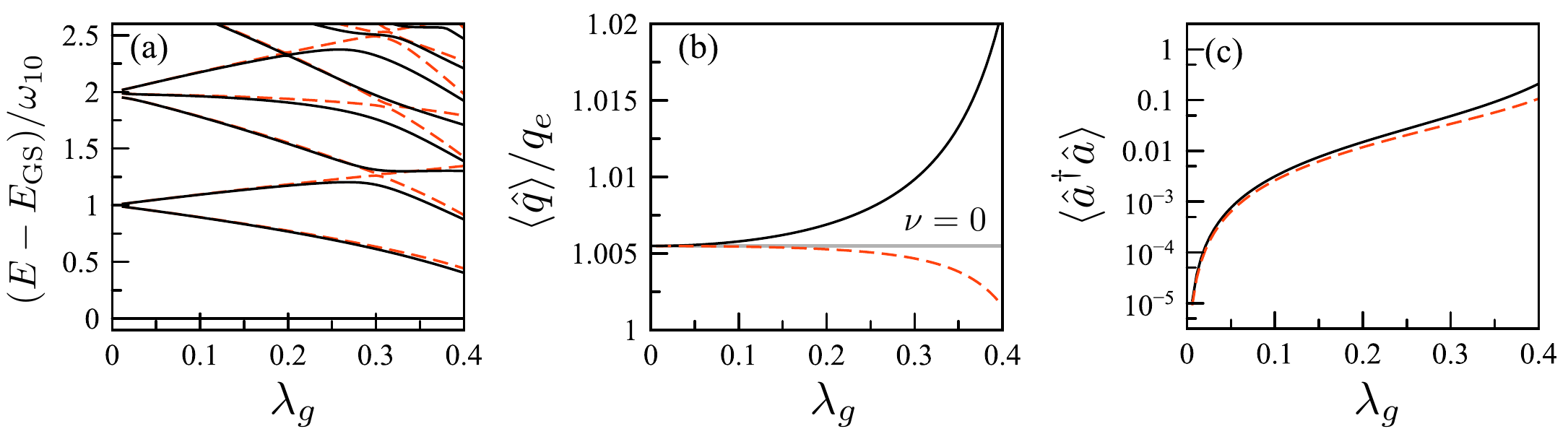}
    \caption{{\bf Static vibrational polariton properties for non-polar molecules}. Polariton spectrum and selected observables as a function of the dimensionless coupling strength $\lambda_g$, for a {\it non-polar} Morse oscillator at equilibrium, in a cavity resonant with the fundamental vibration frequency $\omega_{10}$. In all panels, solid lines are obtained by including transition and permanent dipoles in the Hamiltonian and dashed lines consider transition dipoles only.  (a) Polariton energies relative to the ground state (GS) in units of  $\omega_{10}$.  (b) Mean bond distance $\langle \hat q\rangle$ for the polariton ground state, in units of the equilibrium length $q_e$. The value for the $\nu=0$ vibrational state is shown for comparison. (c) Mean cavity photon number $\langle \hat a^\dagger \hat a \rangle $ for the polariton ground state, in logarithmic scale.}
    \label{fig:spectrum nonpolar}
\end{figure*}

Details about the MCTDH method can be found in Ref. \cite{mctdhpaper}. Briefly, the time-dependent Schr\"{o}dinger equation is solved by introducing the wavepacket ansatz
\begin{equation}    \label{eq:MCTDH ansatz}
    \psi(q,x,t)=\sum_{j_{q}=1}^{n_{q}}\sum_{j_{x}=1}^{n_{x}} A_{j_{q},j_{x}}(t)  \phi_{j_{q}}(q,t) \phi_{j_{x}}(x,t),
\end{equation}
corresponding to an expansion of the system state into a time-varying product basis composed of the single particle function $\phi_{j_{q}}(q,t)$ in the nuclear coordinate and $\phi_{j_{x}}(x,t)$ in the photonic coordinate, each labelled by the integers $j_{q}$ and $j_{x}$, respectively. We set the number of basis functions in the expansion to $n_{q} = n_{x} = 30$. The equations of motion for the coefficients $A_{j_{q},j_{x}}(t)$ and the basis functions on the corresponding 2D product grid are obtained using an open-source MCTDH package \cite{mctdhpack}, and then solved using a numerical integrator. The evolution of system observables $\langle \psi(x,q,t)|\hat O(x,q)|\psi(x,q,t) \rangle$ is obtained from the numerical evaluation of the corresponding integrals on the ($x,q$)-grid.

For static calculations (e.g. polariton spectrum), we use a routine in the MCTDH package that diagonalizes the Hamiltonian matrix on the 2D grid using the Lanczos algorithm \cite{Lanczos}. The dimensionality of the Hamiltonian matrix is $N_{q} N_{x}$, where $N_q$ and $N_x$ are the number of grid points in the nuclear and photonic coordinates, respectively. In the Lanczos method, the eigenvalue problem is transformed into an iterative eigendecomposition. After the number of iterations is set, a defined number of eigenvalues $M$ are computed, with $M<N_{q} N_{x}$.

\section{Results}
\label{sec:results}

\subsection{Static properties of vibrational polaritons}

Our first goal is to compare the static properties of intracavity vibrational polaritons that emerge under strong light-matter coupling for molecular species that are either polar or non-polar at equilibrium (see definition in Table \ref{tab:dipole params}). This comparison is best carried out in the Hilbert space representation because we can controllably neglect the contribution of permanent dipole moments in the light-matter interaction, by setting $\hat{\mathcal{H}_2}=0$ in Eq. (\ref{eq:H2}).

In Fig. \ref{fig:spectrum nonpolar}a we plot the polariton spectrum as a function of the dimensionless light-matter coupling strength $\lambda_g$, for a {\it non-polar} molecule treated in two alternative ways: (\emph{i}) both transition and permanent dipole moments in the vibrational eigenbasis are taken into account (solid curves); (\emph{ii}) only transition dipoles are considered (dashed curves). For a given value of $\lambda_g$, energies are shown relative to the energy of the absolute ground state ($E_{\rm GS}$). The ground level in general exhibits a red shift relative to its energy at $\lambda_g=0$ (not shown). For $\lambda_g\lesssim 0.1$, the lowest excited polariton manifold exhibits the  usual polariton doublet. The second excited manifold has a well-defined triplet structure. At larger coupling strengths $\lambda_g>0.1$, the spectrum develops into several real and avoided crossings between excited polariton levels \cite{Hernandez2019}. If we consider only the contribution of transition dipole moments in the system Hamiltonian (dashed curves), the low-energy excitation spectrum remains mostly unaltered in comparison with the spectrum of the full Hamiltonian for a wide range of coupling strengths. However, for higher coupling strengths $\lambda_g\approx 0.3$, ignoring the permanent dipole moments in the Hamiltonian results in a qualitatively different polariton spectrum.

\begin{figure*}[t]
    \includegraphics[width=0.9\textwidth]{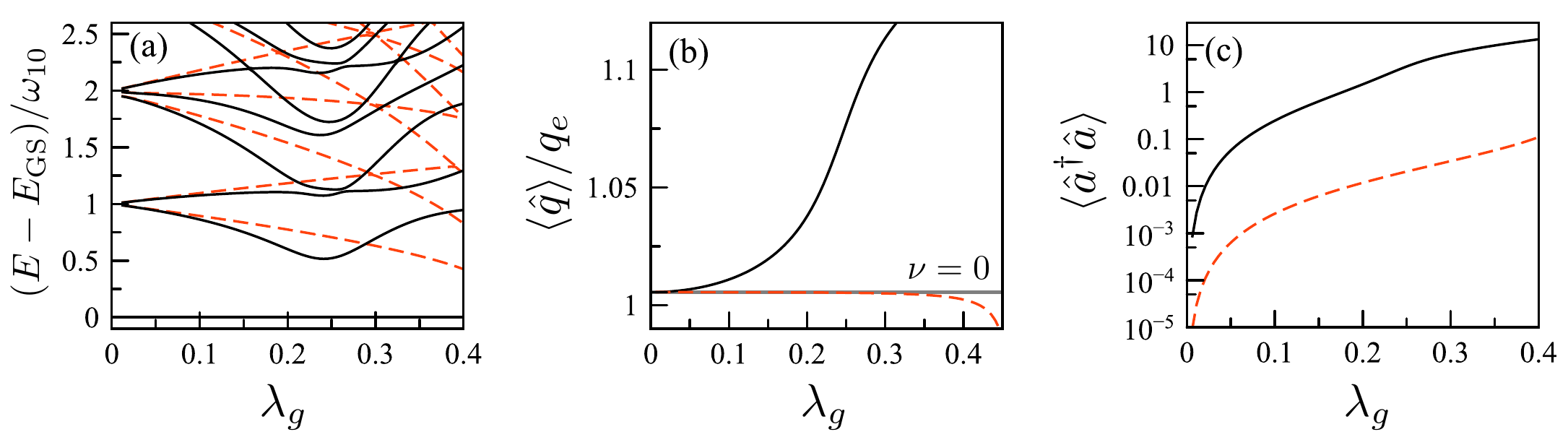}
    \caption{{\bf Static vibrational polariton properties for polar molecules}. Polariton spectrum and selected observables as a function of the dimensionless coupling strength $\lambda_g$, for a {\it polar-right} Morse oscillator at equilibrium, in a cavity resonant with the fundamental vibration frequency $\omega_{10}$. In all panels, the solid lines are obtained by including transition and permanent dipoles in the Hamiltonian and the dashed lines by considering transition dipoles only.  (a) Polariton energies relative to the ground state (GS) in units of  $\omega_{10}$.  (b) Mean bond distance $\langle \hat q\rangle$ for the polariton ground state, in units of the equilibrium length $q_e$. The value for the $\nu=0$ vibrational state is shown for comparison. (c) Mean cavity photon number $\langle \hat a^\dagger \hat a \rangle $ for the polariton ground state, in logarithmic scale.}
    \label{fig:spectrum polar}
\end{figure*}

In panels \ref{fig:spectrum nonpolar}b and \ref{fig:spectrum polar}c, we show the mean bond distance $\langle \hat q\rangle $ and mean intracavity photon number $\langle \hat a^\dagger \hat a\rangle$, respectively, for the  polariton ground state. The model predicts that at moderate values of $\lambda_g\sim 0.1$ and beyond, the system ground state behaves qualitatively different from $\ket{\nu=0}\ket{n_c=0}$, which is the system ground state outside the cavity. This is most evident in panel \ref{fig:spectrum polar}c, where it is shown that the polariton ground state is composed of  Fock states with $n_c\geq 1$ photons, even for coupling strengths as low as $\lambda_g\sim 0.01$. This would be conventionally considered as a signature of ultrastrong light-matter coupling \cite{Kockum2019,Forn-Diaz2018}. 

Panel \ref{fig:spectrum nonpolar}b shows that mean bond distance can be significantly higher in the polariton ground state (solid line) than the bond distance of the $\nu=0$ vibrational eigenstate of the Morse potential considered. This type of {\it mode lengthening} effect is not captured when only transition dipole moments are taken into account in light-matter interaction. By  ignoring the permanent dipole moments of the vibrational states, i.e., setting $\hat{\mathcal{H}}_2=0$, the ground state is predicted to experience bond {\it shortening} (dashed line). This is consistent with the results in Ref. \cite{Hernandez2019}, in which state-dependent permanent dipoles are ignored.

In panel  \ref{fig:spectrum nonpolar}c we show that the intracavity photon number $\langle \hat a^\dagger\hat a\rangle$ has qualitatively the same behavior with the coupling strength whether we take into account permanent dipole moments or not. We discuss below how this seems to only hold for non-polar molecular bonds, as we find substancial differences when we consider vibrational modes that are polar at equilibrium. 

In Fig. \ref{fig:spectrum polar}, we show the polariton spectrum and ground state properties as a function of the coupling strength $\lambda_g$, for nuclear modes that are polar at equilibrium. For concreteness, we consider the {\it polar-right} electric dipole function in Table \ref{tab:dipole params}. In general, the results are more sensitive to the presence or absence of permanent dipole moments in the Hamiltonian than for the case of non-polar molecules. For instance, panel \ref{fig:spectrum polar}a shows that the exact polariton energies in the first and second excited manifolds (solid lines) already differ by a few percent from the energies obtained by neglecting permanent dipole moments (dashed lines). The difference is more evident for values of $\lambda_g\approx 0.2$ and beyond. 

Panel \ref{fig:spectrum polar}b  shows that the {\it mode lengthening} effect already predicted for non-polar molecules in Fig. \ref{fig:spectrum nonpolar} becomes much stronger in the polariton ground state of polar species (solid line). Again, ignoring the contribution of permanent dipole moments predicts a qualitatively different result (dashed line). 

Panel \ref{fig:spectrum polar}c shows that for polar bonds the cavity field can build up a significant amount of photons in the polariton ground state, reaching up to $\langle \hat a^\dagger \hat a\rangle \sim 10$ for $\lambda_g\sim 0.4$ (solid line). Interestingly, this prediction is very sensitive to the presence or absence of permanent dipole moments in the system Hamiltonian. Remarkably, by considering only the contribution of transition dipole moments in the light-matter interaction, the intracavity photon number is consistently underestimated by about two orders of magnitude (dashed line) with respect to the full dipole matrix, for the entire range of coupling strengths considered. This should be compared with Fig. \ref{fig:spectrum nonpolar}c, where only small differences are found.

\subsection{Sub-picosecond polariton dynamics}

Let us now consider the short-time unitary dynamics of the molecular and photon observables $\langle \hat q\rangle $ and $\langle \hat a^\dagger \hat a\rangle$, for an intracavity system initially prepared in a polariton wavepacket of the form
\begin{equation}\label{eq:initial wavepacket}
 \ket{\psi_0} = \sum_\nu\sum_{n_c} \beta_{\nu\,n_c}\ket{\nu}\ket{n_c},
\end{equation}
where $\beta_{\nu \,n_c}$ are complex wavepacket coefficients in the diabatic basis $\ket{\nu}\ket{n_c}$. Transforming Eq. (\ref{eq:initial wavepacket}) to the coordinate space representation preserves the values of  $\beta_{\nu \,n_c}$. The initial state is then propagated either in Hilbert space or in coordinate space as described in Sec. \ref{sec:propagation methods}.

\begin{figure}[t]
    \includegraphics[width=0.45\textwidth]{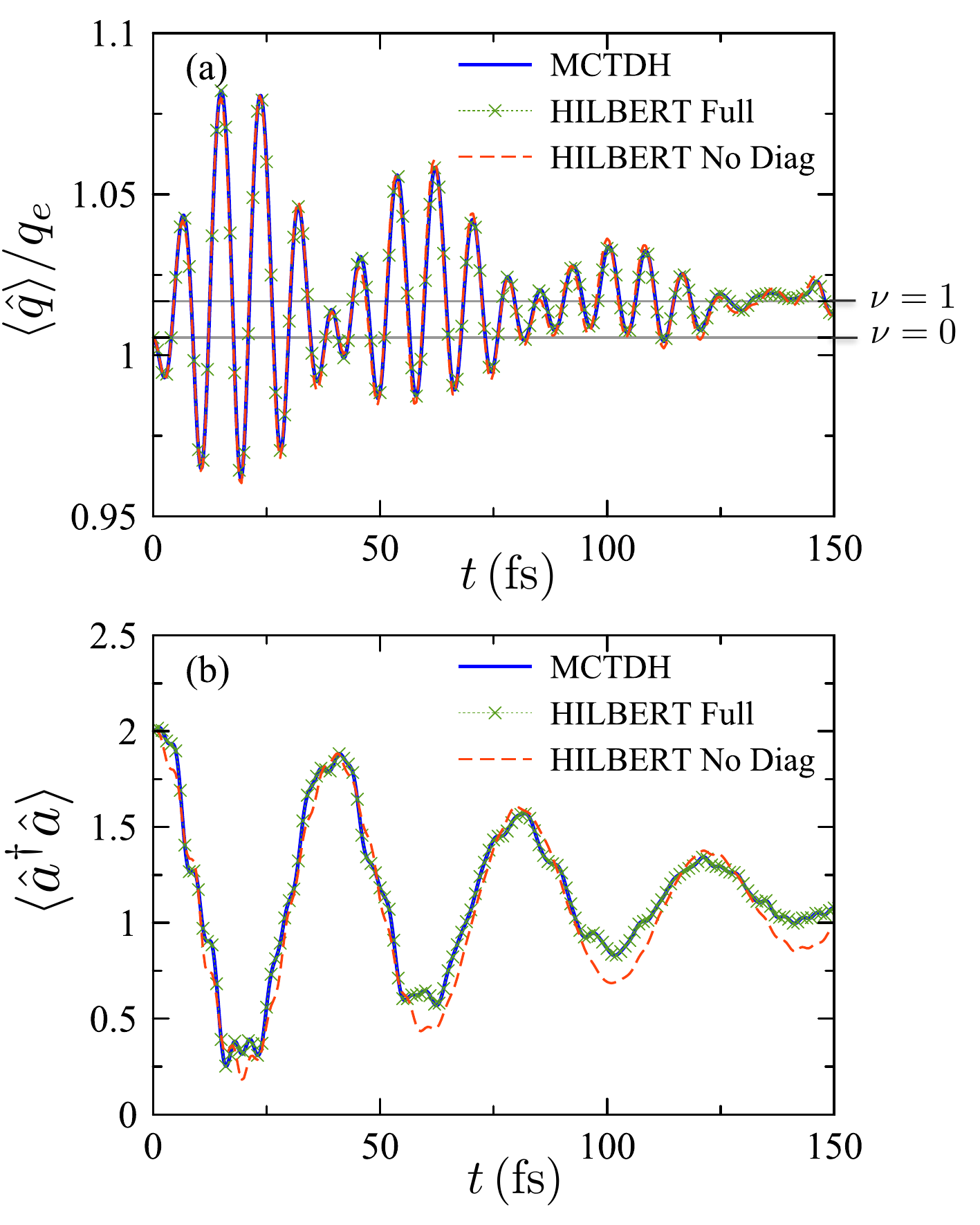}
    \caption{{\bf Polariton wavepacket evolution for non-polar molecular vibrations}. (a) mean mode length $\langle \hat q\rangle$ as a function of time. As a reference, horizontal lines show the static values of the $\nu=0$ and $\nu=1$ vibrational levels outside the cavity. (b) Mean intracavity photon number $\langle \hat a^\dagger \hat a\rangle$. Results are shown as obtained with the MCTDH method (solid lines), the Hilbert space method with the full dipole matrix (crosses), and the Hilbert space method without diagonal (permanent) dipole matrix elements (dashed). In both panels, the molecular vibration is set to be initially in its $\nu=0$ vibrational level and the cavity in a coherent state with $|\alpha|^2=2$ photons. The cavity frequency is resonant with the fundamental vibrational frequency $\omega_{10}$, and the light-matter coupling strength is $\lambda_g=0.125$.}
    \label{fig:evolution coherent}
\end{figure}

\subsubsection{Non-polar molecular vibrations}

For concreteness, we first consider an initial condition that describes a system in which an individual molecular vibration in its ground level ($\nu=0$), is embedded at $t=0$ into a cavity that was previously prepared in a coherent state (e.g., via laser pumping). The molecular vibration is assumed to be non-polar at equilibrium (see Table \ref{tab:dipole params}), and the cavity initially has $|\alpha|^2={2}$ photons, with $\alpha$ being the coherent state amplitude. We set $\lambda_g=0.125$. 

Physical intuition suggests that a molecule could be expected to interact strongly with the intracavity photons of the resonant cavity and become vibrationally excited, absorbing cavity photons. This intuitive picture is reproduced by converged MCTDH calculations in Figure \ref{fig:evolution coherent}.  Over the first 150 fs, the molecule is seen to increase its mode length from the bare value at $\nu=0$ to the value for $\nu=1$, by absorbing a single cavity photon on average.

\begin{figure}[t]
    \includegraphics[width=0.45\textwidth]{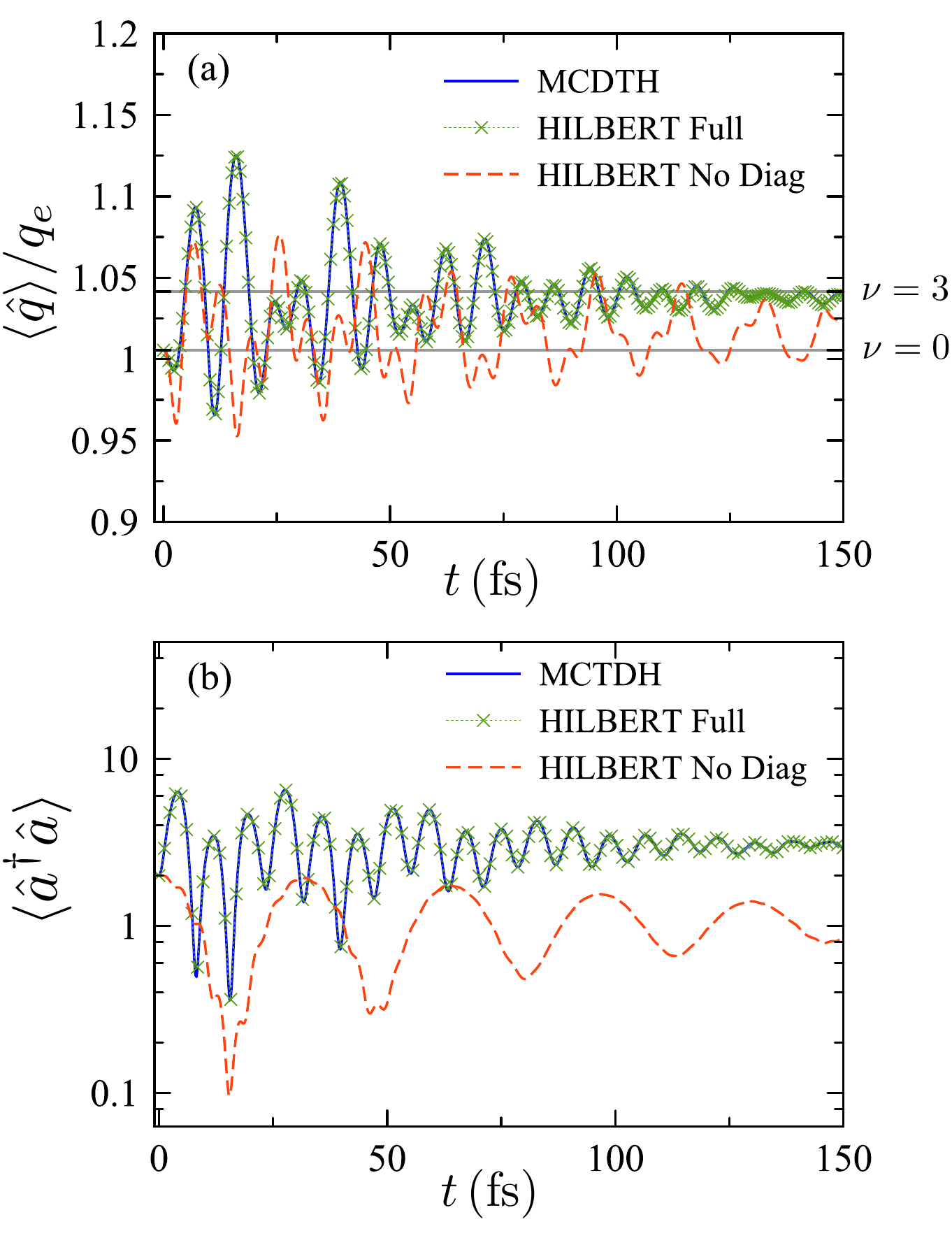}
    \caption{{\bf Polariton wavepacket evolution for polar molecular bonds}. (a) mean mode length $\langle \hat q\rangle$ as a function of time, for a {\it polar-right} vibration. Horizontal lines correspond to the mode lengths of the $\nu=0$ and $\nu=3$ vibrational levels outside the cavity. (b) Mean intracavity photon number $\langle \hat a^\dagger \hat a\rangle$. Results are shown as obtained with the MCTDH method (solid lines), the Hilbert space method with the full dipole matrix (crosses), and the Hilbert space method without diagonal (permanent) dipole matrix elements (dashed). In both panels, the molecular vibration is set to be initially in its $\nu=0$ vibrational level and the cavity in a coherent state with $|\alpha|^2=2$ photons. The cavity frequency is resonant with the fundamental vibrational frequency $\omega_{10}$, and the light-matter coupling strength is $\lambda_g=0.125$.}
    \label{fig:evolution coherent polar}
\end{figure}

This behavior is also seen on timescales on the order of the vibrational period ($2\pi/\omega_{10}=8.27$ fs). Over the first 40 fs, nuclear oscillations that result in an overall mode lengthening (panel \ref{fig:evolution coherent}a) are accompanied by a step-wise decrease of the cavity photon number (panel \ref{fig:evolution coherent}b). The partial recurrence of the photon number toward its initial value in turn results in the overall shortening of the vibrational mode. This qualitative behavior then tends to repeat in cycles, but the amplitudes of the oscillations in the nuclear and photonic observables become reduced at later times. 

Figure \ref{fig:evolution coherent} also shows that evolution of the polariton observables  using the Hilbert space approach gives results that are indistinguishable from those obtained using the MCTDH method, as long as the entire dipole matrix is used, i.e.,  both transition dipoles (off-diagonal) and permanent dipoles (diagonal) contribute to light-matter coupling. Small deviations from the exact MCTDH evolution are found when we neglect the permanent dipole moments, which is most evident in the evolution of the photon number (panel \ref{fig:evolution coherent}b). 

\subsubsection{Polar molecular vibrations}

We now consider the polariton dynamics of vibrational modes that are polar at equilibrium. More specifically, we focus on {\it dipole-right}  species, as defined in Table \ref{tab:dipole params}. We assume the same initial condition as in the previous case, i.e., the vibration initially in $\nu=0$ and the cavity field in a coherent state with real amplitude $\alpha=\sqrt{2}$. In addition to comparing the qualitative differences in the system dynamics with respect to the case of non-polar species, we again compare the results obtained using a numerically exact MCTDH approach with those obtained using a Hilbert space approach that either includes or neglects the contribution of permanent dipole moments in the light-matter interaction. 

Figure \ref{fig:evolution coherent polar} shows the resulting evolution of the mean mode length and the intracavity photon number for a {\it dipole-right} molecule. The coupling strength $\lambda_g$ is slightly higher than in Fig. \ref{fig:evolution coherent}. Despite this difference, the comparison of Figs. \ref{fig:evolution coherent} and  \ref{fig:evolution coherent polar} suggests that the intuitive physical picture of a vibrational mode overall being excited by absorbing a cavity photon is not universal. For polar species, MCTDH calculations predict an overall increase of {\it both} the mean mode length and cavity photon number over the first 150 fs, as the result of light-matter interaction (solid lines). Quantitatively equivalent results are obtained by computing the dynamics in Hilbert space with both diagonal and off-diagonal dipole matrix elements taken into account (cross symbols). When we ignore the contribution of permanent dipoles in the Hamiltonian (dashed lines), we obtain results that are qualitatively different. Specifically, both the mean photon number and mode length are consistently underestimated if permanent dipoles are neglected. 

Let us first discuss in detail the evolution of the mean photon number (panel \ref{fig:evolution coherent polar}b). Over short timescales on the order of the bare vibrational period ($8.27$ fs), the system exhibits alternating patterns of mode lengthening at the expense of photon absorption and mode shortening accompanied by re-emission of photons into the cavity field. These cycles were also found to occur for non-polar species in Fig. \ref{fig:evolution coherent}. However, for polar bonds our model predicts a qualitatively new physical behavior: The light-matter system rapidly develops a sizable number of cavity photons over the first few vibrational periods. More specifically, in less than 5 fs the cavity field amplitude rapidly grows from its initial condition to  about $\langle \hat a^\dagger \hat a\rangle \approx 6.3$ photons. Subsequent evolution of the photon number occurs in cycles of decreasing amplitude over the first 100 fs, until the polariton wavepacket tends to stabilize into a cavity with more photons than the initial state. 

Let us now consider the evolution of the mean mode length (panel \ref{fig:evolution coherent polar}a). Qualitatively, we obtain a similar behavior to the case of non-polar molecules (Fig. \ref{fig:evolution coherent}), i.e., the molecule experiences an overall mode lengthening in comparison with its initial ground state configuration ($\nu=0$). However, the molecular bond is lengthened by the cavity field to a greater extent for polar species, at equal coupling strength ($\lambda_g=0.125$). For the system in panel \ref{fig:evolution coherent polar}a, the mode length reaches a quasi-steady value that is comparable with the length of a bare Morse vibrational state with $\nu=3$ vibrational quanta.

\subsection{Non-classical intracavity initial conditions}

\begin{figure}[t]
    \includegraphics[width=0.44\textwidth]{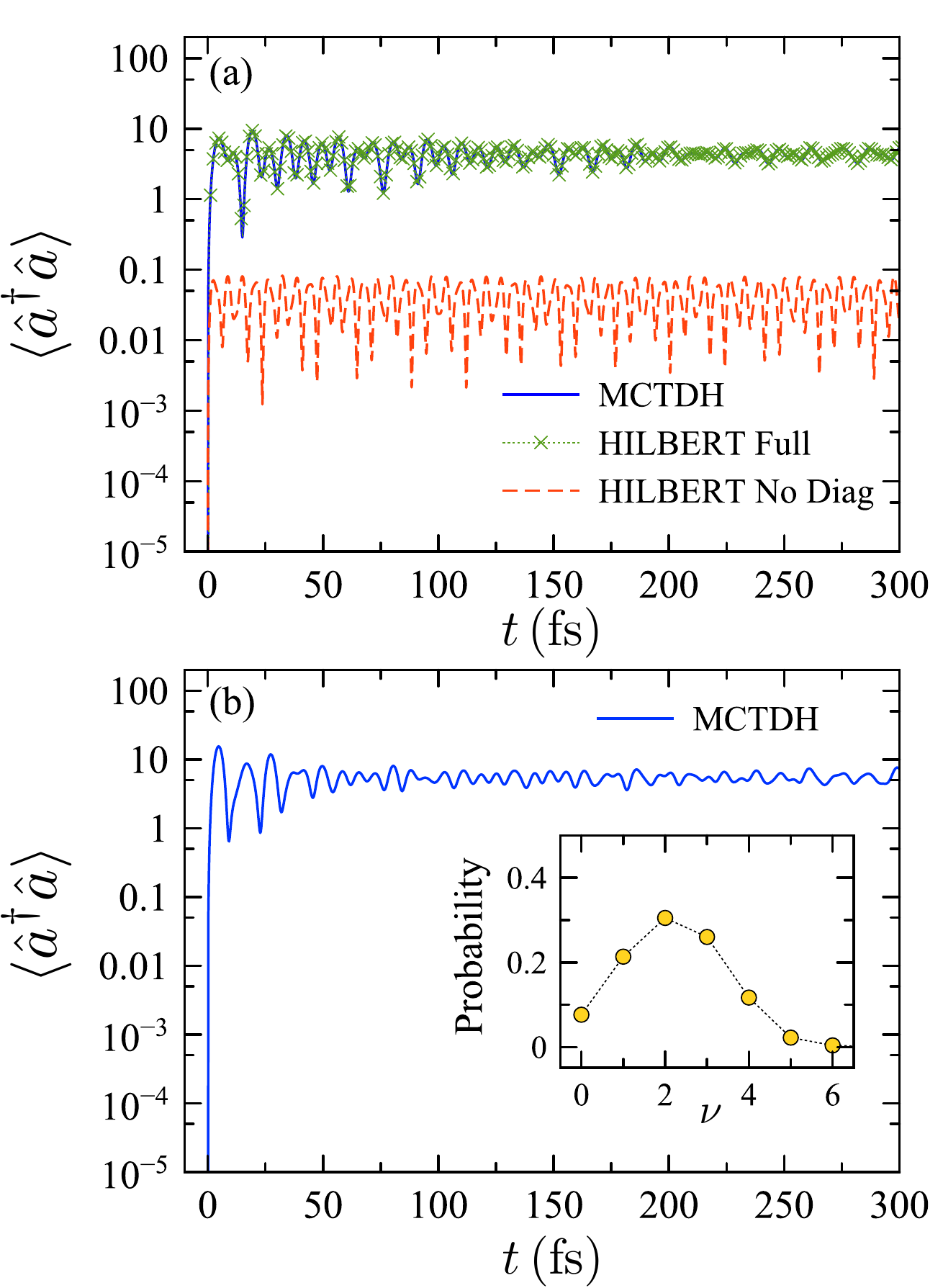}
    \caption{{\bf Polariton evolution for polar vibrations in a vacuum field}. Mean photon number $\langle \hat a^\dagger \hat a\rangle$ as a function of time, for a {\it polar-right} molecular species, for two initial molecular conditions: (a) vibrational ground state $\ket{\nu=0}$. (b) Vibrational wavepacket $\ket{\psi_{\rm v}}=\sum_\nu c_\nu \ket{\nu}$, with the vibrational distribution $|c_\nu|^2$ shown in the inset. Results are shown as obtained with the MCTDH method (solid line), the Hilbert space method with the full dipole matrix (crosses), and the Hilbert space method without diagonal (permanent) dipole matrix elements (dashed line). In both panels the cavity is initially in the vacuum state $\ket{n_c=0}$. The cavity frequency is resonant with the fundamental vibrational frequency $\omega_{10}$, and the light-matter coupling strength is $\lambda_g=0.25$.}
    \label{fig:evolution ground vacuum}
\end{figure}

Coherent states are known to provide an accurate description of lasers and are thus considered to have classical field statistics \cite{Gerry2005}. Let us consider now a qualitatively different scenario where the cavity field is prepared initially in a state that has no classical analogue. We are interested in understanding the differences that can be expected in the dynamics of vibrational polariton when the cavity field evolve from a pure Fock state, relative to the results in Figs. \ref{fig:evolution coherent} and \ref{fig:evolution coherent polar}, which correspond to a cavity in an initial coherent state. If there are qualitative differences, we also want to understand the role of the polarity of the molecular bond in the evolution of the polariton system. 

We first consider a light-matter system in which a polar molecular vibration is first prepared in its ground state and then adiabatically embedded into a cavity in its vacuum state. Such a preparation scheme may be feasible with microfluidic infrared cavities \cite{George2015}. The initial wavepacket in the diabatic basis is simply 
\begin{equation}\label{eq:00 initial state}
\ket{\psi_0}\equiv\ket{\nu=0}\ket{n_c=0}.
\end{equation} 
In Fig. \ref{fig:evolution ground vacuum}, we show the evolution of the mean intracavity photon number starting from the initial state in Eq. (\ref{eq:00 initial state}), for a {\it polar-right} molecular vibration (see definition in Table \ref{tab:dipole params}), for the coupling strength parameter $\lambda_g=0.25$. The numerically exact MCTDH results  (panel \ref{fig:evolution ground vacuum}a, solid line) show that starting from the exact vacuum at $t=0$, the cavity field develops a significant amplitude over the first few femtoseconds, with the photon number reaching up to $\langle \hat a^\dagger \hat a\rangle\approx 7.3$ in 5 fs. The polariton wavepacket then evolves into a state with a photon number that undergoes moderate fluctuations about a relatively large number (e.g., $\langle \hat a^\dagger \hat a\rangle\approx 4.6$ in 300 fs). Calculations performed with the Hilbert space approach taking into account both transition and permanent dipole moments (crosses), give results that are equivalent to those obtained with MCTDH.  However, when we ignore the contribution of permanent dipole moments in the light-matter interaction (dashed line), the results consistently underestimate the intracavity photon number by at least two orders of magnitude over all timescales considered.

For comparison purposes, in panel \ref{fig:evolution ground vacuum}b we show the evolution of mean intracavity photon number starting from a cavity in the vacuum Fock state, but the molecule in a vibrational wavepacket, i.e., $\ket{\psi_0}=(\sum_\nu c_\nu\ket{\nu})\otimes\ket{n_c=0}$. The wavepacket has a mean vibrational energy $\bar E_{\rm vib}=2.126 \omega_{10} $ and is initialized at equilibrium ($\langle q\rangle_0=q_e$). The overall photon number evolution is qualitatively similar to the case in panel \ref{fig:evolution ground vacuum}a, suggesting that photon generation is more related to the structure of the light-matter coupling Hamiltonian than the degree of initial molecular excitation.  

We also perform similar calculations using non-polar molecules initialized in the diabatic state $\ket{\nu=0}\ket{n_c=0}$ (not shown). For the same light-matter coupling parameters as in Fig. \ref{fig:evolution ground vacuum}, we find that for this class of molecules, the mean mode length does not significantly vary in comparison with free space, and also the average number of cavity photons that can be produced remains at least two orders of magnitude smaller than the case of polar molecules, even after several hundred femtoseconds. 

\subsection{Role of the slope of the dipole function}

Up to this point, we have numerically established that the sub-picosecond dynamics of vibrational polaritons formed with polar vibrations is qualitatively different from the evolution of polaritons formed by vibrations that are non-polar at equilibrium (e.g., compare Figs. \ref{fig:evolution coherent} and \ref{fig:evolution coherent polar}). We now discuss what appears to be unexpected feature of vibrational polaritons for polar molecules: not only the value of the dipole moment at equilibrium is relevant to determine the dynamics of vibrational polaritons, but equally important is the {\it sign} of the slope of the electric dipole function. 

We support this conjecture below by comparing the dynamics of polariton observables for {\it polar-right} and {\it polar-left} vibrational modes (see definitions in Table \ref{tab:dipole params}). In our Morse oscillator model, the transition dipole moments ($d_{\nu'\nu}$) of {\it polar-left} and {\it polar-right} vibrations have the same qualitative structure as a function of the vibrational quantum number $\nu$ (see Fig. \ref{fig:dipoles}), but their permanent dipole moments ($d_{\nu}$) behave qualitatively different. $d_{\nu}$ decreases monotonically with $\nu$ for polar-left vibrations. For {\it polar-right} species,  $d_{\nu}$ increases with $\nu$ for low quantum numbers, then decays to small values for higher vibrational levels. In other words, the discrete slope $\Delta d_\nu/\Delta \nu$ has opposite signs at low $\nu$, for our two types of polar vibrations. 

\begin{figure}[t]
    \includegraphics[width=0.45\textwidth]{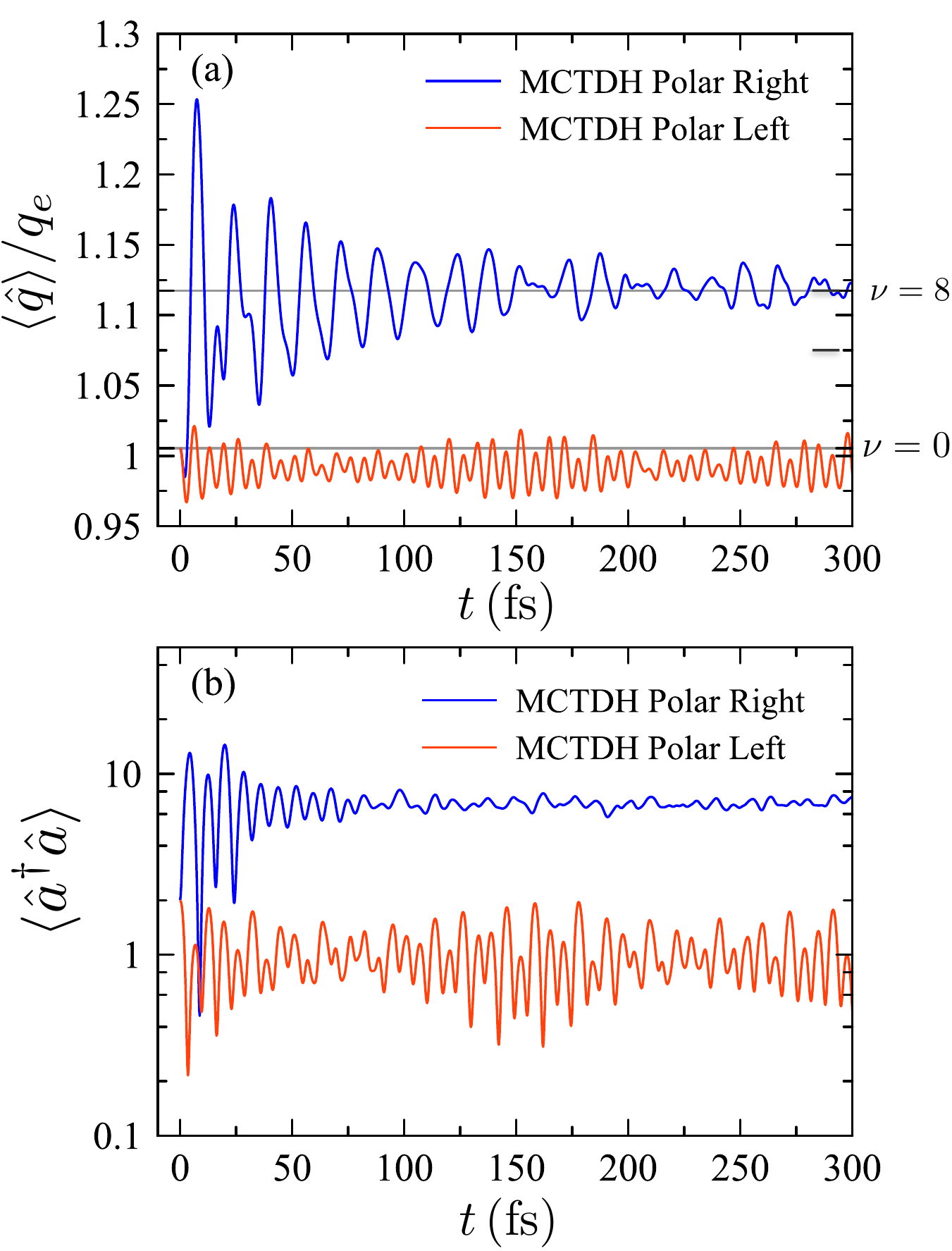}
    \caption{{\bf Polariton evolution for polar vibrations with different dipole function slopes at equilibrium}. (a) mean mode length $\langle \hat q\rangle$ as a function of time, for a {\it polar-right} vibration (blue line) and {\it polar-left} vibration (red line). Horizontal lines correspond to the mode lengths of the $\nu=0$ and $\nu=8$ vibrational levels outside the cavity. (b) Mean intracavity photon number $\langle \hat a^\dagger \hat a\rangle$. In both panels, the molecular vibration is set to be initially in its $\nu=0$ vibrational level and the cavity in a coherent state with $|\alpha|^2=2$ photons. Results are obtained with the MCTDH method. The cavity frequency is resonant with the fundamental vibrational frequency $\omega_{10}$, and the light-matter coupling strength is $\lambda_g=0.25$.}
    \label{fig:left right comparison}
\end{figure}

\begin{figure*}[t]
    \includegraphics[width=0.9\textwidth]{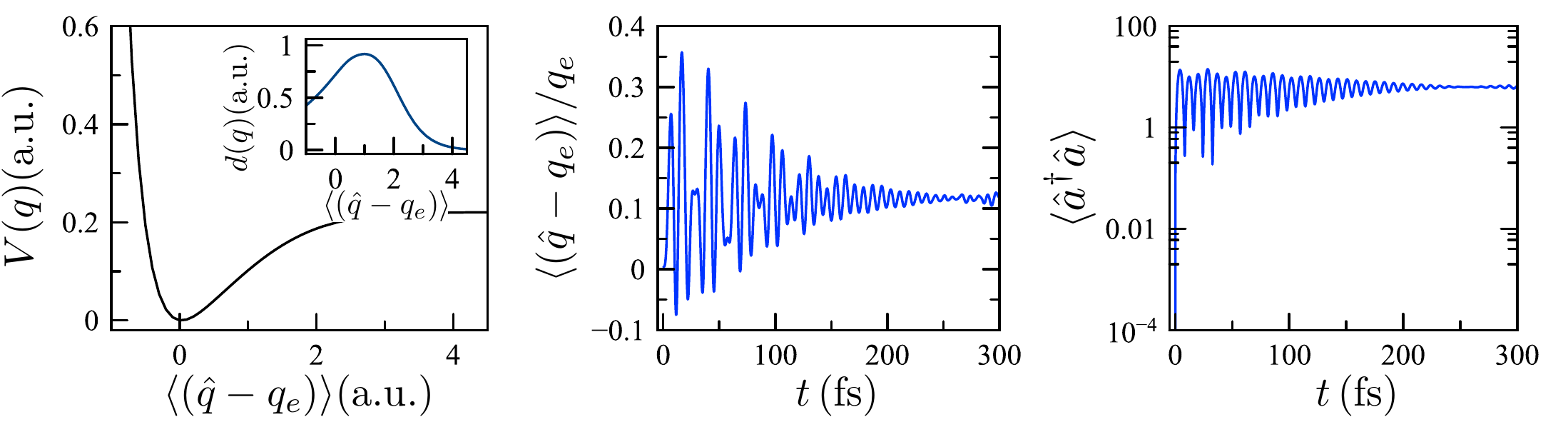}
    \caption{{\bf Vibrational polariton evolution for the asymmetric stretching mode of hydrogen fluoride}.(a) {\it ab-initio} potential energy curve along the vibrational coordinate. The inset shows the corresponding electric dipole function around the equilibrium configuration $q_e$. (b) Fractional bond displacement from equilibrium as a function of time. (c) Mean intracavity photon number. The molecule is initially in the vibrational ground state and the cavity is in the vacuum state. The cavity frequency is resonant with the fundamental vibrational frequency $\omega_{10}$, and the light-matter coupling strength is $\lambda_g=0.1$.}
    \label{fig:HF evolution}
\end{figure*}

In Fig. \ref{fig:left right comparison}, we correlate the sign of the slope $\Delta d_\nu/\Delta \nu$ for {\it polar-left} and {\it polar-right} species, with the evolution of the mean mode length and intracavity photon number under strong coupling. The polariton dynamics is computed using MCTDH. For simplicity, we consider the same initial condition as in Fig. \ref{fig:evolution coherent polar}, i.e., molecules are initially in their ground vibrational level and the cavity is in a coherent state with two photons on average. 

Panel \ref{fig:left right comparison}a shows that starting from the same initial condition, polar-right anharmonic vibrations experience an overall increase of the mode length over the first few hundred femtoseconds, while polar-left species effective decrease their mode length in the same time interval. For a coupling strength $\lambda_g=0.25$, the mode length of a polar-right molecule increases to a value comparable with a $\nu=8$ vibrational eigenstate. On the contrary, a polar-left vibrational mode can only transiently exceed the initial mode length of the $\nu=0$ vibrational level.

Panel \ref{fig:left right comparison}b in turn shows that the lengthening of the vibrational mode for a polar-right species is accompanied by a significant increase of the intracavity photon number. For the parameters considered, the cavity evolves from the vacuum state into a polariton wavepacket with $\langle \hat a^\dagger \hat a\rangle\approx 7.0$ in about 300 fs. In contrast, for a polar-left species the cavity on average {\it gives away} about one photon on average into the material system. 

\subsection{Case examples: CO$_2$ and HF molecules}

\begin{figure*}[t]
    \includegraphics[width=0.9\textwidth]{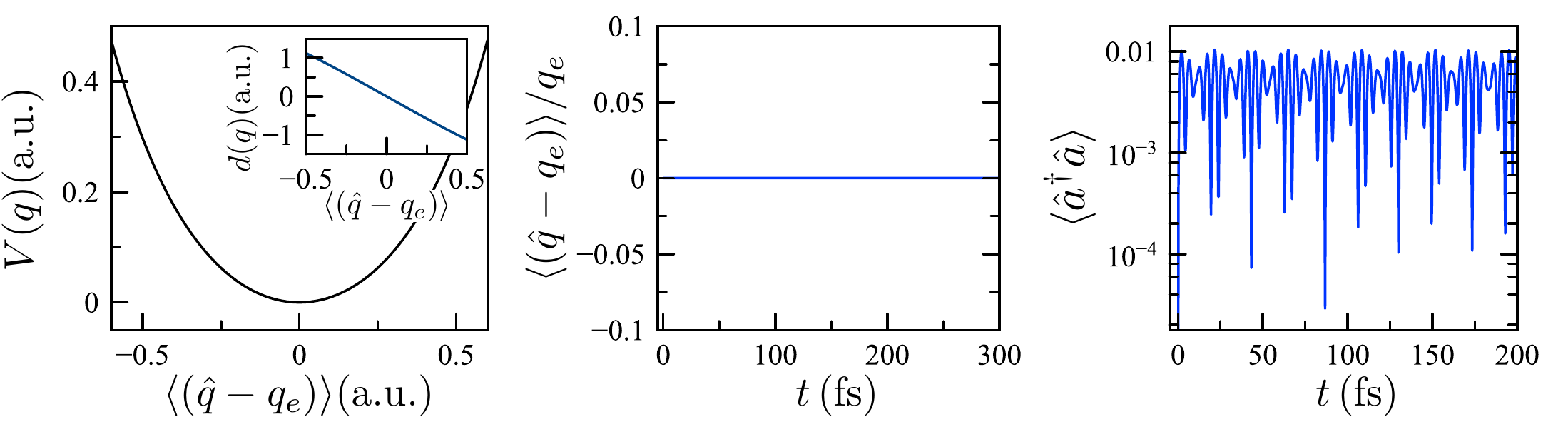}
    \caption{{\bf Vibrational polariton evolution for the asymmetric CO stretching mode of carbon dioxide}. (a) {\it ab-initio} potential energy curve along the vibrational coordinate near the equilibrium configuration $q_e$. The inset shows the corresponding electric dipole function. (b) Fractional bond displacement from equilibrium as a function of time. (c) Mean intracavity photon number. The molecule is initially in the vibrational ground state and the cavity is in the vacuum state. The cavity frequency is resonant with the fundamental vibrational frequency $\omega_{10}$, and the light-matter coupling strength is $\lambda_g=0.1$.}
    \label{fig:CO2 evolution}
\end{figure*}

Until now we have considered the dynamics of idealized Morse vibrations under strong light-matter coupling. Although Morse oscillators are reasonably good approximations to the stretching modes of many molecular species, so far we have not shown that the light-matter physics predicted above can be observed with real molecules. 

Let us  consider the polariton dynamics of an individual hydrogen fluoride (HF) molecule in an infrared cavity resonant with its fundamental frequency ($\tilde \nu_{10} = 3990$ cm$^{-1}$, $2\pi/\omega_{10}=8.36$ fs). In Fig. \ref{fig:HF evolution}a, we plot the {\it ab-initio} potential energy curve $V(q)$ and  electric dipole function $d(q)$ for the ground electronic state ($X^1\Sigma^+$). These were  obtained using the electronic structure package MOLPRO. In both cases, a complete active space (CAS) calculation of the lowest electronic state  has been performed using a multiconfigurational self-consistent field (MCSCF) method, then followed by multireference configuration interaction (MRCI) method, using Dunning's correlation consistent basis set with diffuse functions aug-cc-pVQZ. According to our nomenclature, HF belongs to the {\it polar-right} class of molecules. 

In panels Fig. \ref{fig:HF evolution}b and \ref{fig:HF evolution}c, we show the evolution of the mean mode length $\langle \hat q\rangle $ and the intracavity photon number $\langle \hat a^\dagger \hat a\rangle $, respectively, for a vibrational polariton wavepacket that evolves from an HF molecule initially prepared in its ground vibrational state ($\nu=0$) with a cavity in a vacuum Fock state ($n_c=0$). The results show that the molecule experiences an overall increase in its mode length of about 10\% in 300 fs, accompanied by a significant buildup of intracavity photons. This behavior is consistent with the results in Fig. \ref{fig:left right comparison}  for {\it polar-right} anharmonic oscillators, as expected. 

In a second example, let us consider the dynamics of an individual carbon dioxide molecule (CO$_2$) in an infrared cavity resonant with the asymmetric stretching mode ($\tilde \nu_{10}=3860.7$ cm$^{-1}$, $2\pi/\omega_{10}=8.64$ fs). This vibrational mode is non-polar at equilibrium but acquires a dipole moment away from it, thus belonging to the {\it non-polar} class. In Fig. \ref{fig:CO2 evolution}a, we plot the {\it ab-initio} ground state potential energy curve ($^1A'$ state) and electric dipole function near equilibrium along the asymmetric stretching mode. These were obtained using the same {\it ab-initio} method used for HF (MCSCF/aug-cc-pVQZ). 

In panels \ref{fig:CO2 evolution}b and \ref{fig:CO2 evolution}c, we show the evolution of the polariton mode length and photon number, respectively, over the first few hundred femtoseconds. The polariton wavepacket is also assumed to evolve from a vibrationless molecule in a vacuum Fock state. Unlike the cases considered in previous sections, the mode length in CO$_2$ is found to remain invariant at its bare equilibrium value throughout the evolution of the polariton state, which we attribute to the negligible anharmonicity of the potential energy curve. In contrast, the intracavity photon number does varies significantly from its initially value zero, rapidly reaching up to $\langle \hat a^\dagger \hat a\rangle\sim 10^{-3}$ after a few vibrational periods, from where it undergoes quasi-stationary oscillations. 

\section{Discussion and Outlook}
\label{sec:discussion}

We have shown in this work that the resonant interaction of an individual  molecular vibration with a quantized cavity field can have very different physical observable consequences, depending on the dipolar properties of the molecular electron density. In principle, this is not completely unexpected if we recall that the oscillator strength $S_{\nu'\nu}$ of infrared absorption band for a vibrational transition $\nu\rightarrow \nu'$ is proportional to the square of the slope of the electric dipole function $d_e(q)$ at the equilibrium configuration \cite{Demtroder-book}. 

Our results show that under conditions of strong and ultrastrong light-matter interaction, not only the slope but the {\it entire shape} of the electric dipole function is important to understand the static and dynamical properties of vibrational polaritons. For instance, we find that two polar molecules which would have nominally identical infrared absorption bands, can undergo qualitatively different polariton dynamics under resonant coupling with a quantized cavity field. 

Consider the case of {\it polar-right} and {\it non-polar} vibrations, following the nomenclature in Fig. \ref{fig:dipoles} and Table \ref{tab:dipole params}. Since they have essentially the same value of $|d_{10}|$, the two types of molecular vibrations would give the same infrared absorption spectrum, provided they also have the same fundamental frequency $\omega_{10}$. However, the evolution of physical observables for these two types of vibrations are qualitatively different when they are placed inside a resonant cavity. For instance, our results show that by preparing a {\it polar-right} vibration in its ground vibrational state outside the cavity, then place it suddenly into the cavity volume in total darkness (no cavity photons), the system evolves into a hybrid light-matter wavepacket that behaves as if the molecular mode lengthens over sub-picosecond timescales, while the cavity spontaneously generates photons in the same time interval (see Fig. \ref{fig:evolution ground vacuum}). 

In another unexpected result, we find that polar molecular vibrations can either experience bond-lengthening or bond-shortening, depending on the shape of their electric dipole function (see Fig. \ref{fig:left right comparison} for comparison of {\it polar-right} and {\it polar-left} species). For interaction strengths at the conventional onset of  ultrastrong coupling \cite{Kockum2019,Forn-Diaz2018}, we find that {\it polar-right} vibrations  can increase its length by up to 10\% from its equilibrium value (see Fig. \ref{fig:HF evolution} for hydrogen fluoride), while developing a intracavity field with up to about $10$ photons on average. 

\begin{figure*}[t]
    \includegraphics[width=0.9\textwidth]{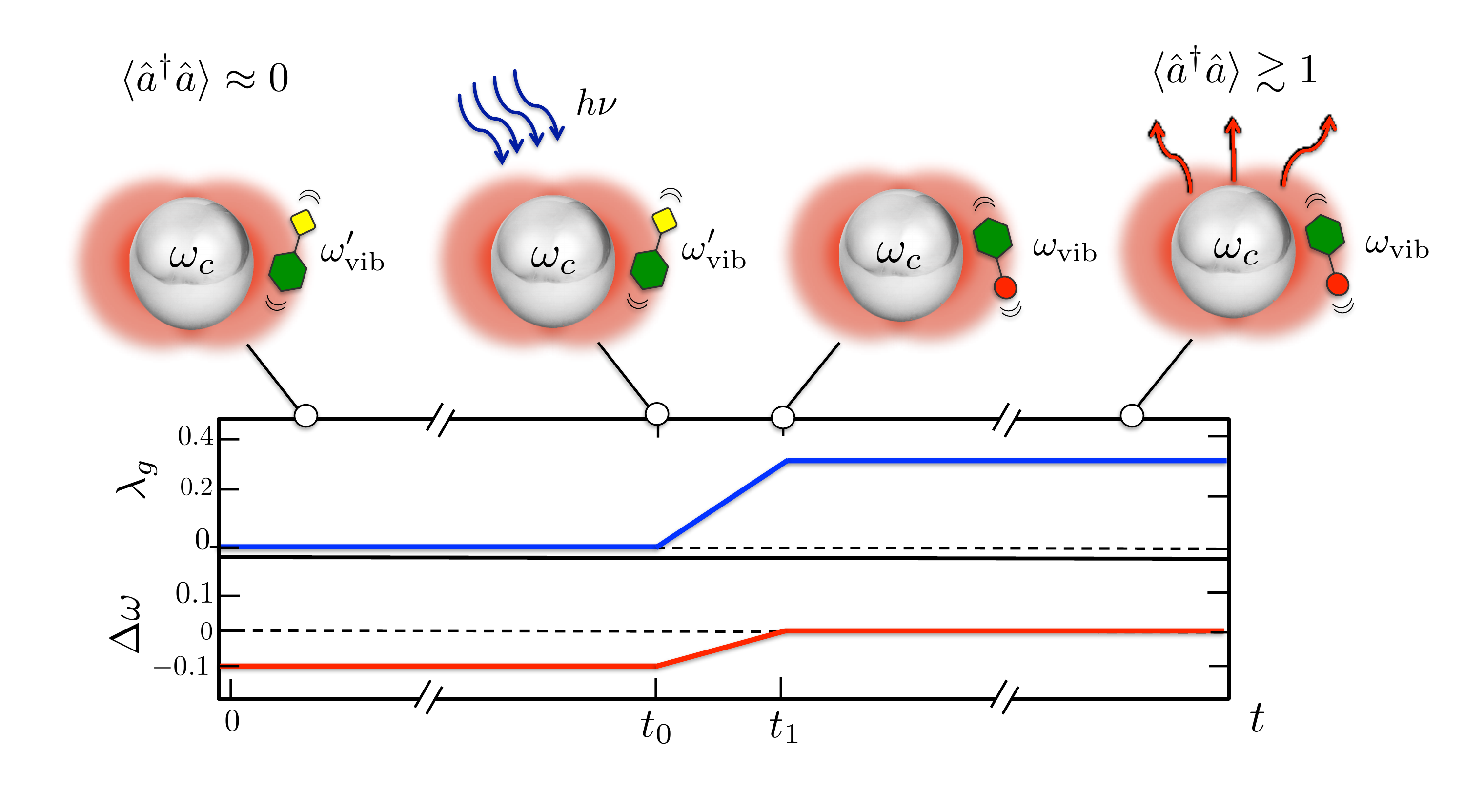}
    \caption{{\bf Proposed scheme for the generation of near-field infrared quantum light using photochemistry}. An individual vibrating molecule adsorbed on a plasmonic nanoparticle undergoes a photochemical reaction at $t_0$ over the timescale $T_{\rm chem}= |t_1-t_0|$, producing a molecular bond that couples strongly with the near-field vacuum of the nanostructure. Depending on the duration of the chemical transformation, the light-matter system is prepared at $t_1$ into a vibrational polariton wavepacket that can emit quantum light into the far field through radiative relaxation. The panel at the bottom qualitative describes the associated time evolution of the light-matter coupling strength $\lambda_g$, and the detuning $\Delta \omega$ between the relevant plasmonic frequency $\omega_{\rm c}$ and the vibrational frequency $\omega_{\rm vib}$.}
    \label{fig:preparation scheme}
\end{figure*}

Mode lengthening in a strongly coupled infrared cavity may in turn result in different chemical reactivities of polar molecules, in comparison with free space. Our theoretical and numerical analysis may thus provide a consistent  basis for the development of a reaction rate theory for vibrational polaritons that can offer a microscopic understanding of the observed chemical reaction enhancements under vibrational strong coupling \cite{George2015}, which remains to be an outstanding goal in the field. 

For polar and non-polar molecular vibrations we predict an ultrafast dynamical buildup of intracavity photons during the evolution of a strongly coupled light-matter system with initially no vibrational or photonic excitations. This spontaneous generation of photons is a natural outcome of wavepacket evolution in the polariton energy basis, since diabatic initial product states $\ket{\nu}\ket{n_c}$ are not eigenstates of the light-matter Hamiltonian for any finite coupling strength ($\lambda_g\neq 0$). In the adiabatic polariton basis, vibration-photon product states can thus be seen as wavepackets that have broad distribution polariton energies. As long as the wavepacket has contributions from polariton excited states, then the system can radiate intracavity photons into the far field through radiative decay.

The energy needed to produce cavity field excitations together with the vibrational excitation of a molecule comes from light-matter interaction Hamiltonian itself. This is simpler to visualize in the Hilbert space representation [see Eqs. (\ref{eq:H1})-(\ref{eq:H2})]. The counter-rotating terms of light-matter term can directly couple diabatic states $\ket{\nu}\ket{n_c}$ that can be interpreted as a simultaneous excitation of both the vibrational mode and the cavity field. If the contribution of these counter-rotating couplings is significant, we can thus observe an overall increase of the photon number and mode length. 

In experiments, it may be challenging to initialize a strongly coupled cavity-vibration system in a diabatic product state $\ket{\nu}\ket{n_c}$, with a definite number of vibrational excitations and a definite photon number. One potential preparation scheme is illustrated in Fig. \ref{fig:preparation scheme}. At early times ($t\ll t_0)$, an individual molecule is placed within a relevant plasmonic mode volume in total darkness (i.e., no laser driving). Although the vacuum field amplitude at the molecular-nanoparticle interface can be large, the molecular vibration does not exchange energy with the near-field  vacuum because either the fundamental frequency $\omega_{\rm vib}'$ is far detuned from the relevant plasmon frequency $\omega_{\rm cav}$ ($\Delta \omega\equiv  \omega_c-\omega_{\rm vib}'\neq 0$), or the dipole moment for the relevant vibrational transition is such that light-matter coupling is weak ($\lambda_g\approx 0$), or both conditions occur simultaneously (as in Fig. \ref{fig:preparation scheme}). Under these conditions, the molecule-nanoparticle system will simply thermalize with its environment and remain unaltered in the absence of additional external perturbations. In particular, the number of photons in the near field will not exceed the level imposed by the background radiation at infrared frequencies, which is negligibly small ($\langle \hat a^\dagger \hat a\rangle\approx 0$). 

Strong light-matter coupling is suddenly activated over the time interval $T_{\rm chem}= t_1-t_0$ by chemically converting the adsorbed molecule into a species with a vibrational mode that either becomes resonant with the near-field vacuum ($\Delta\omega\approx 0$), or the relevant vibrational transition dipole is such that light-matter coupling becomes strong ($\lambda_0\gtrsim 0.1$), or both situations occur simultaneously (as in Fig. \ref{fig:preparation scheme}). UV-Vis photochemistry \cite{Zhang2019} or electron tunneling \cite{Hla2000} can be used to activate the bond-forming reaction on the surface of the nanoparticle at $t_0$ . This chemical transformation is not expected to form infrared near-field photons. After the reaction is complete ($t\geq t_1$), the molecule-nanoparticle system is left in the strong coupling regime of light-matter interaction. Assuming that the strongly coupled bond is formed in the vibrational ground state, the system at $t_1$ is left in the diabatic product state $\ket{\nu=0}\ket{n_c=0}$, from where it evolves into a polariton wavepacket that can eventually radiate a number of infrared photons into the far field over sub-picosecond timescales ($\langle \hat a^\dagger \hat a\rangle\gtrsim 1$). In the absence of external infrared driving fields, the generation of near-field quantum light stops when the wavepacket relaxes to the vibrational polariton ground state. 

In this proposed preparation scheme, the chemical reaction timescale $T_{\rm chem}$ is very important. If it is comparable or much shorter than the Rabi oscillations expected to establish for the vibrational populations under strong light-matter coupling, then system evolves as described in the main text. If $T_{\rm chem}$ is much larger than the relevant Rabi period, then the light-matter system evolves {\it adiabatically} from an uncoupled product eigenstate $\ket{\nu}\ket{n_c}$ into a polariton eigenstate, not a polariton wavepacket. For instance, if the uncoupled eigenstate is the bare vibrationless vacuum $\ket{\nu=0}\ket{n_c=0}$, then the system is prepared {adiabatically}  in the polariton ground state at $t_1$, from where no photons can be emitted. Radiation can only be emitted as long as the system wavefunction $\ket{\psi(t=t_1)}$ has finite contributions from excited polariton states. 

In summary, we have developed the first detailed theoretical framework for understanding the sub-picosecond  dynamics of anharmonic vibrational polaritons. Starting from a fundamental light-matter interaction model in the electric dipole approximation, we correlate the dynamics of both material and photonic observables of a strongly coupled cavity-vibration system with the underlying electrostatic properties of vibrations determined by the molecular electronic structure. Using numerically exact quantum dynamics methods, we show that the entire shape of the electric dipole function of anharmonic vibrational modes is relevant for understanding the  sub-picosecond dynamics of vibrational polaritons in infrared cavities. Our single-molecule analysis may stimulate further developments on vibrational strong coupling with nanophotonics \cite{Luxmoore2014,Metzger2019,Ojambati2019}. The results can also be extended in order to take into account many-body effects, as well as photonic or material dissipative processes. Such extensions would enable a more direct comparison with available experimental evidence in liquid-phase cavities \cite{George2015,Dunkelberger2016,Dunkelberger2019,Imran2019}

\vspace{4cm}

\section{Acknowledgments}

J.T. is supported by CONICYT through the Postdoctoral Fellowship Grant No. 3200565.  F.H. is supported by CONICYT through grants REDES ETAPA INICIAL, Convocatoria 2017 No. REDI 170423, and FONDECYT Regular No. 1181743. All authors thank the generous support of Iniciativa Cient\'ifica Milenio (ICM) through the Millennium Institute for Research in Optics (MIRO).

\bibliography{biblio}
\bibliographystyle{unsrt}

\end{document}